# Graphene terahertz modulators by ionic liquid gating

*Yang Wu, Chan La-o-vorakiat, Xuepeng Qiu, Jingbo Liu, Praveen Deorani, Karan Banerjee, Jaesung Son, Yuanfu Chen, Elbert E. M. Chia,\* and Hyunsoo Yang\**

Y. Wu, Prof. H. Yang
Department of Electrical and Computer Engineering, Graphene Research Centre, National University of Singapore, 117576 Singapore
Centre for Advanced 2D Materials and Graphene Research Centre, National University of Singapore, 6 Science Drive 2, 117546 Singapore
E-mail: eleyang@nus.edu.sg

Dr. C. La-o-vorakiat
Faculty of Science, King Mongkut's University of Technology Thonburi, Bangkok 10140, Thailand
Division of Physics and Applied Physics, School of Physical and Mathematical Sciences, Nanyang Technological University, Singapore 637371, Singapore

Dr. X. Qiu, P. Deorani, K. Banerjee, Dr. J. Son
Department of Electrical and Computer Engineering, National University of Singapore, 117576 Singapore

J. Liu, Prof. Y. Chen
State Key Laboratory of Electronic Thin Films and Integrated Devices, University of Electronic Science and Technology of China, Chengdu 610054, China

Prof. E. Chia
Centre for Advanced 2D Materials and Graphene Research Centre, National University of Singapore, 6 Science Drive 2, 117546 Singapore
Division of Physics and Applied Physics, School of Physical and Mathematical Sciences, Nanyang Technological University, Singapore 637371, Singapore
E-mail: elbertchia@ntu.edu.sg

Electromagnetic waves cover the range from X-ray to radio frequencies. Within this spectrum, the utilization of terahertz (THz) wave, ranging from 0.1 THz ($10^{12}$ Hz) to 30 THz, was lagging until 1980s.[1-3] Recently, many interesting properties of THz wave have been discovered, which suggests promising applications, such as spectroscopy, safety surveillance, cancer diagnosis, imaging, and communication.[1, 4, 5] However, THz applications are limited due to the lack of high performance optical components, for example sources,[4] detectors,[6] phase modulators,[7, 8] and intensity modulators.[9-11] In particular, intensity modulators in the THz regime operate mainly by controlling the carrier density in semiconductors.[12, 13]

Nevertheless, due to the high insertion losses, the intensity of THz signal is greatly attenuated.[14] Vanadium oxide ($VO_2$) is an alternative choice for modulating THz intensity via insulator-to-metal phase transitions at 340 K.[15] However, conventional temperature control using $VO_2$ is not convenient in integrated systems. Recently, THz modulation was realized by electrical gating on $VO_2$ and the modulation effect was originated from electrochemical modification, which indicates promising applications of $VO_2$ as THz modulators.[16] In contrast to all these devices above, graphene based devices offer outstanding performance with great electrical controllability.[17, 18]

Graphene, a single atomic layer of carbon, has attracted huge interest in a wide range of studies. Graphene has been recognized for excellent mechanical strength, chemical stability, and the highest electrical mobility of carriers due to the unique conical band structure.[19, 20] These characteristics enable the application of graphene in modulators in both visible[21, 22] and THz range[10, 11, 17, 23, 24]. The modulation in the visible range is enabled by interband transitions limiting the absorption to only 2.3%.[25, 26] On the other hand, in the THz range, intraband transitions of graphene dominate and the electric-field amplitude modulation is much more significant. A total (100%) electric-field modulation of a graphene based THz modulator was predicted, but the modulation depth was demonstrated to be 15% experimentally.[10] Furthermore, it was theoretically demonstrated that graphene based THz modulators could have very low insertion losses by optimizing the substrates.[10] Nevertheless, further improvements in the modulation depth are required for practical applications.

In this work, we experimentally demonstrate and numerically support the excellent performance of THz modulators based on graphene/ionic liquid/graphene sandwich structures. The modulation covers a broadband frequency range from 0.1 to 2.5 THz with the modulation depth of up to 99% by applying a small gate voltage of 3 V. To our knowledge, this is the highest modulation ratio from graphene based THz devices to date. The outstanding performance of graphene based device benefits from two key components: (1) a linear conical band structure of graphene and (2) a powerful gating effect of ionic liquid. First, due to the linear band structure of graphene, the Fermi level in the vicinity of the Dirac point can be linearly controlled by tuning the gate voltage, which subsequently changes the transmittance. Second, the strong gating effect of ionic liquid derives from the fact that charges accumulate within several nanometers in proximity to the graphene/ionic liquid interfaces.[27, 28] Consequently, the magnitude of electric fields on graphene is very large, which results in the effective tuning of graphene's Fermi level. For the massless charge carriers in graphene, there is power-law dependence $|E_F| \propto |n|^{1/2}$ in between the Fermi level and the carrier concentration.[29] With this relationship, it can be known that the gating is essentially tuning the carrier concentration in graphene (or electrical conductivity).[30] According to the Drude model, graphene optical conductivity is equivalent to its electrical conductivity at THz regime.[10, 31] As a result, the electrical gating is capable for modulating THz waves. Moreover, our sandwich structures make use of the high mobility of both electrons and holes,[20] unlike the previous semiconductor based modulators,[12, 13] where the hole contributions were negligible.

The proposed THz modulators are based on a sandwich structure of two layers of graphene and ionic liquid (**Figure 1**a). When a voltage is applied between two graphene films, holes are accumulated on one of the graphene films, while electrons are on the other side (Figure 1b). The experimental results from a single-layer graphene based modulator are presented in **Figure 2**. By increasing the applied voltage from 0 to 3 V, the electric field amplitude of the transmitted THz pulses decreases significantly (Figure 2a). In contrast, we observe that no gate-dependent phase change is introduced to the THz pulses, and the time shift in Figure 2a is added for clarity. The insertion loss in this device is 48%, which is mainly introduced by the quartz glasses and graphene films. The insertion losses could be reduced by engineering the quartz glass thickness[10] and reducing the defects in CVD graphene films." THz traces are truncated between two dashed lines in the inset of Figure 2b to remove the effects due to multiple reflections, and then the spectra of transmitted THz pulses (Figure 2b) are obtained using a fast Fourier transformation (FFT). The strength of THz electric fields attenuates across the observed THz range as a function of the gate voltage. The spectra in Figure 2c are normalized to the spectrum with zero gate voltage, and the overall THz electric field decreases as the gate voltage increases. The oscillating features in the normalized spectra arise from the multi-reflection of THz wave inside the cavity between two graphene films (Supporting Information Section 1). The normalized THz transmittance, $T = (t_v/t_0)^2$, where $t_v$ is the electric field strength of transmitted THz at gate voltage $V_g$, and $t_0$ is at $V_g = 0$. By averaging the THz spectra across the entire detected bandwidth (0.1 to 2.5 THz), the transmittance decreases monotonically with respect to the gate voltage (Figure 2d). We observe a power modulation depth $M = (1-T)$ of up to 83% with an applied voltage of 3 V, in which each layer of graphene contributes to 60% modulation. The modulation depth is significantly greater than the previously reported value (15%) from a single-layer graphene based device.[10] We also test the repeatability and reliability of our devices by alternately switching the gate voltage among 0 and ±3 V (Figure 2d, inset). The maximum applied gate voltage is restricted up to 3 V, which is under the electrochemical window of the graphene films.[32]

In order to have a better insight into the device operation, we perform a series of studies to understand the role of the Fermi level change[8, 14] and the carrier redistribution within the ionic liquid. The first control study investigates the importance of the interface between the ionic liquid and graphene. As illustrated in **Figure 3**a, a thin (~0.5 μm) non-conductive photoresist (ma-N2405) layer is coated for isolating the ionic liquid from graphene. As the capacitance of ionic liquid is three orders larger than that of photoresist, majority gate voltage is loaded onto the ionic liquid. If the carrier distribution change in ionic liquid is the origin of the THz intensity modulation, the modulation level in this control device should be the same as the original ones in Figure 2. However, the peak values of these THz pulse have a negligible change with gate voltage even up to 20 V (Figure 3b), and the time domain signal also remains consistent, as shown in the inset of Figure 3b (horizontally shifted for clarity). The second control study was conducted with two different cell thicknesses of 50 and 100 μm (Figure 3c). No significant difference in the device performances is observed with only a slight variation in modulation percentage. The above two control studies suggest that (1) the interfaces between the ionic liquid and graphene play an important role for the THz modulation and (2) the contribution from the bulk ionic liquid is negligible. Therefore, we conclude that the THz modulation is caused by manipulating the Fermi level of graphene due to the electric fields inside the graphene layers.

In addition, similar devices based on bi-layer and tri-layer graphene (Figure 3d) are studied. Since the multi-layer graphene was prepared by stacking single-layer graphene without interlayer atomic bonds, it is expected that the graphene layer at the graphene/ionic liquid interface mostly contributes to the THz modulation. Therefore, devices with different graphene thicknesses should show a similar modulation to the single-layer graphene based device.[22] In Figure 3d, however, the modulation depths at 3 V show 83%, 89% and 93% for single-layer, bi-layer, and tri-layer graphene based devices, respectively. This differences can be understood by the fact that the overall domain boundary defects in CVD graphene are eliminated in a multilayer graphene configuration.[33] We also demonstrate a further enhancement of the modulation depth to 99% by simply stacking two devices (a bi-layer and a tri-layer graphene device) on top of each other (Figure 3d). The different trends in the data curves, such as the modulation depth differences and the peak position shift, are due to the graphene quality variation between films, which is induced during the growth and the transfer processes. For applications, the optimal numbers of graphene layers and stacked devices should be chosen considering the trade-off among the device conductivity, durability, and insertion loss.[8]

As shown in **Figure 4**a, the THz transmittance modulation is symmetric under positive and negative gate voltages (Supporting Information, Figure S2), as expected for devices with such a symmetric structure. The voltage dependent transmittance can be understood by three states of band alignment (Figure 4b) of the top (blue) and bottom (green) graphene layers. Without any gate voltage, the Fermi level of both graphene layers is slightly lower than the Dirac point due to unintentional p-dopants.[34, 35] For a large gate voltage of $V_g \gg V_{Dirac}$, the Fermi level of the bottom graphene film increases, while that of the top film decreases. As a result, the total density of states (DOS) at the Fermi level of both graphene layers increases, leading to the enhancement of THz absorption and reflection due to greater intraband transition. This mechanism also explains the symmetric modulation with respect to the polarity of the gate voltage, when the field direction is reversed.

Small peaks are observed at low applied voltages indicated by red arrows in Figure 4a, corresponding to the gate voltage of $V_g \approx \pm 2V_{Dirac}$. When a small gate voltage is applied, the Fermi level of the bottom graphene layer rises and approaches the Dirac point (decreasing the DOS), while the Fermi level of top layer falls (increasing the DOS). At this gate voltage regime, the DOS from graphene layers compensate each other, giving rise to a saturation behavior in the transmittance. The saturation takes place in the range of $-2V_{Dirac} < V_g < +2V_{Dirac}$ when the doping levels of both graphene layers are equal and the band dispersion of graphene is linear. However, a slight increase of the transmittance in Figure 4a is observed. The existence of the peaks in transmittance suggests that the positive and negative charges accumulate on graphene films unequally. There are a few reasons for this phenomenon. First of all, when considering impurity induced scatterings, different degrees of correlations for electrons and holes contribute to the asymmetric gating effect.[36, 37] The dominant Coulomb scattering at the vicinity of Dirac point due to ionized impurities can also introduce a non-linearity in graphene's band structure.[38] It is also reported that the gating effect is more effective for the anions than that of cations in the ionic liquid.[22] In addition, the grain boundaries in CVD graphene play an important role for the electrical property.[39, 40]

In order to further understand the proposed THz modulators, we simulate the absorption, transmission, and reflection by the transfer matrix method.[41] Detailed calculations are shown in the Supporting Information (Section 3).[10, 14] The conductivity of the top and bottom graphene layer was measured by the four-probe method (Figure 5a) in tri-layer graphene based devices, and the sheet conductivity of graphene is assumed to be equivalent to its DC conductivity.[14] From Figure 5a, it can be seen that the gating capability of ionic liquid on graphene is roughly 1 mS/V. Using the conductivity change per gate voltage, we calculate the transmittance, absorption, and reflectance with respect to different gate voltages (Figure 5b). The modulation trend in Figure 5b agrees well with the THz measurement in Figure 4a. Interestingly, the simulated reflection also increases with the gate voltage, suggesting that the devices can operate in the reflection geometry as well. We can infer the increment of reflectivity as function of voltage by inspecting the oscillating feature of the transmittance spectra in Figure 2c. As the gate voltage increases, the stronger internal reflections from the graphene layers are detected, which is in line with the results in Figure 5b.

The proposed THz modulators by ionic liquid gating on graphene layers benefit from the excellent modulation depth, low operation voltage, and low insertion losses, in contrast to the previous semiconductor-based modulators. By tuning the Fermi level of the top and bottom graphene layers, we modulate the THz transmittance by up to 93% from a single modulator device and up to 99% by stacking two devices. We further confirm the modulation mechanism by calculating the transmittance using the transfer matrix method. The proposed device provides a new platform to explore high performance THz devices using graphene and ionic liquids.

**Experimental Section**

*Device preparation:* The proposed THz modulators are based on a sandwich structure of two layers of graphene and ionic liquid (Figure 1a). Cr (10 nm)/Au (40 nm) electrodes (1 mm by 10 mm in scale) are thermally evaporated onto the graphene substrate in a vacuum level of $5\times10^{-8}$ Torr. Then two quartz glass substrates with graphene films on top are sandwiched with two plastic spacers in between. The gap between the graphene films is determined by the thickness of spacers (50 or 100 μm). The quartz glasses are slightly misaligned to avoid any short circuit between top and bottom layers. 1-ethyl-3-methylimidazolium bis (tri-fluoromethylsulfonyl) imide ([EMIM][TFSI]) ionic liquid is injected into the cavity among the spacers and graphene films. Since the electric fields near the two graphene films are in opposite directions under a gate voltage, holes are accumulated on one of the graphene films, while electrons are on the other side (Figure 1b). Because the electrical resistance of graphene is much smaller than that of ionic liquid, the electric field is evenly distributed in the structures.

*Graphene preparation:* graphene films are prepared by the chemical vapor deposition (CVD) method.[42] Single-layer graphene is grown on a copper foil, and then transferred to the quartz glasses. Multilayer graphene films are prepared by stacking multiple single-layer graphene films. The Raman spectroscopy results show prominent G and 2D peaks, while the D peak is negligible, indicating a good quality of graphene films.[8, 35, 43] Because there are no atomic bonds between the different layers in our multilayer graphene, each stacked layer has consistent chemical and optoelectrical properties similar to single-layer graphene.[31]

*THz measurements:* For transmittance measurements, the THz pulses are generated by a photoconductive antenna based time domain spectroscopy system (Supporting Information, Figure S5), providing a bandwidth from 0.1 to 2.5 THz. The acquired data are averaged from 900 spectra with the scanning frequency of 30 Hz, which gives a very high signal to noise ratio.

**Acknowledgements**

This research was supported by NRF-CRP "Novel 2D materials with tailored properties: beyond graphene" (No. R-144-000-295-281), NRF-CRP "Ultrafast study of oxide interface" (No. NRF-CRP4-2008-04), and MOE AcRF Tier 1 (RG 13/12 and MOE2014-T1-001-056).

[1]     B. Ferguson, X. C. Zhang, *Nat. Mater.* **2002**, 1, 26.
[2]     M. Tonouchi, *Nat. Photonics* **2007**, 1, 97.
[3]     S. L. Dexheimer, *Terahertz spectroscopy: principles and applications*, CRC press, Danvers **2007**.
[4]     P. H. Siegel, *IEEE Trans. Microwave Theory Tech.* **2002**, 50, 910.
[5]     W. Withayachumnankul, G. M. Png, Y. Xiaoxia, S. Atakaramians, I. Jones, L. Hungyen, B. Ung, J. Balakrishnan, B. W. H. Ng, B. Ferguson, S. P. Mickan, B. M. Fischer, D. Abbott, *Proc. IEEE* **2007**, 95, 1528.
[6]     L. Vicarelli, M. S. Vitiello, D. Coquillat, A. Lombardo, A. C. Ferrari, W. Knap, M. Polini, V. Pellegrini, A. Tredicucci, *Nat. Mater.* **2012**, 11, 865.
[7]     C.-Y. Chen, T.-R. Tsai, C.-L. Pan, R.-P. Pan, *Appl. Phys. Lett.* **2003**, 83, 4497.
[8]     Y. Wu, X. Ruan, C.-H. Chen, Y. J. Shin, Y. Lee, J. Niu, J. Liu, Y. Chen, K.-L. Yang, X. Zhang, J.-H. Ahn, H. Yang, *Opt. Express* **2013**, 21, 21395.
[9]     Z. Xie, X. Wang, J. Ye, S. Feng, W. Sun, T. Akalin, Y. Zhang, *Sci. Rep.* **2013**, 3, 3347.
[10]    B. Sensale-Rodriguez, R. Yan, M. M. Kelly, T. Fang, K. Tahy, W. S. Hwang, D. Jena, L. Liu, H. G. Xing, *Nat. Commun.* **2012**, 3, 780.
[11]    P. Weis, J. L. Garcia-Pomar, M. Höh, B. Reinhard, A. Brodyanski, M. Rahm, *ACS Nano* **2012**, 6, 9118.
[12]    D. Shrekenhamer, S. Rout, A. C. Strikwerda, C. Bingham, R. D. Averitt, S. Sonkusale, W. J. Padilla, *Opt. Express* **2011**, 19, 9968.
[13]    T. Kleine-Ostmann, K. Pierz, G. Hein, P. Dawson, M. Marso, M. Koch, *J. Appl. Phys.* **2009**, 105, 093707.
[14]    B. Sensale-Rodriguez, T. Fang, R. Yan, M. M. Kelly, D. Jena, L. Liu, H. Xing, *Appl. Phys. Lett.* **2011**, 99, 113104.
[15]    F. Fan, Y. Hou, Z.-W. Jiang, X.-H. Wang, S.-J. Chang, *Appl. Opt.* **2012**, 51, 4589.
[16]    M. D. Goldflam, M. K. Liu, B. C. Chapler, H. T. Stinson, A. J. Sternbach, A. S. McLeod, J. D. Zhang, K. Geng, M. Royal, B.-J. Kim, R. D. Averitt, N. M. Jokerst, D. R. Smith, H.-T. Kim, D. N. Basov, *Appl. Phys. Lett.* **2014**, 105, 041117.
[17]    B. Sensale-Rodriguez, R. Yan, S. Rafique, M. Zhu, W. Li, X. Liang, D. Gundlach, V. Protasenko, M. M. Kelly, D. Jena, L. Liu, H. G. Xing, *Nano Lett.* **2012**, 12, 4518.
[18]    R. Degl'Innocenti, D. S. Jessop, Y. D. Shah, J. Sibik, J. A. Zeitler, P. R. Kidambi, S. Hofmann, H. E. Beere, D. A. Ritchie, *ACS Nano* **2014**, 8, 2548.
[19]    A. K. Geim, K. S. Novoselov, *Nat. Mater.* **2007**, 6, 183.
[20]    A. H. Castro Neto, F. Guinea, N. M. R. Peres, K. S. Novoselov, A. K. Geim, *Rev. Mod. Phys.* **2009**, 81, 109.
[21]    M. Liu, X. Yin, E. Ulin-Avila, B. Geng, T. Zentgraf, L. Ju, F. Wang, X. Zhang, *Nature* **2011**, 474, 64.
[22]    E. O. Polat, C. Kocabas, *Nano Lett.* **2013**, 13, 5851.
[23]    L. Ren, Q. Zhang, J. Yao, Z. Sun, R. Kaneko, Z. Yan, S. Nanot, Z. Jin, I. Kawayama, M. Tonouchi, J. M. Tour, J. Kono, *Nano Lett.* **2012**, 12, 3711.
[24]    S. H. Lee, M. Choi, T.-T. Kim, S. Lee, M. Liu, X. Yin, H. K. Choi, S. S. Lee, C.-G. Choi, S.-Y. Choi, X. Zhang, B. Min, *Nat. Mater.* **2012**, 11, 936.

[25]    R. R. Nair, P. Blake, A. N. Grigorenko, K. S. Novoselov, T. J. Booth, T. Stauber, N. M. R. Peres, A. K. Geim, *Science* **2008**, 320, 1308.
[26]    J. Niu, V. G. Truong, H. Huang, S. Tripathy, C. Qiu, A. T. S. Wee, T. Yu, H. Yang, *Appl. Phys. Lett.* **2012**, 100, 191601.
[27]    S. Ono, K. Miwa, S. Seki, J. Takeya, *Appl. Phys. Lett.* **2009**, 94, 063301.
[28]    J. Son, K. Banerjee, M. Brahlek, N. Koirala, S.-K. Lee, J.-H. Ahn, S. Oh, H. Yang, *Appl. Phys. Lett.* **2013**, 103, 213114.
[29]    L. Ju, B. Geng, J. Horng, C. Girit, M. Martin, Z. Hao, H. A. Bechtel, X. Liang, A. Zettl, Y. R. Shen, F. Wang, *Nat. Nanotechnol.* **2011**, 6, 630.
[30]    D. N. Basov, R. D. Averitt, D. van der Marel, M. Dressel, K. Haule, *Rev. Mod. Phys.* **2011**, 83, 471.
[31]    X. Zou, J. Shang, J. Leaw, Z. Luo, L. Luo, C. La-o-vorakiat, L. Cheng, S. A. Cheong, H. Su, J.-X. Zhu, Y. Liu, K. P. Loh, A. H. Castro Neto, T. Yu, E. E. M. Chia, *Phys. Rev. Lett.* **2013**, 110, 067401.
[32]    T. Y. Kim, H. W. Lee, M. Stoller, D. R. Dreyer, C. W. Bielawski, R. S. Ruoff, K. S. Suh, *ACS Nano* **2010**, 5, 436.
[33]    X. Li, C. W. Magnuson, A. Venugopal, J. An, J. W. Suk, B. Han, M. Borysiak, W. Cai, A. Velamakanni, Y. Zhu, L. Fu, E. M. Vogel, E. Voelkl, L. Colombo, R. S. Ruoff, *Nano Lett.* **2010**, 10, 4328.
[34]    K. S. Novoselov, A. K. Geim, S. V. Morozov, D. Jiang, Y. Zhang, S. V. Dubonos, I. V. Grigorieva, A. A. Firsov, *Science* **2004**, 306, 666.
[35]    Y. J. Shin, G. Kalon, J. Son, J. H. Kwon, J. Niu, C. S. Bhatia, G. Liang, H. Yang, *Appl. Phys. Lett.* **2010**, 97, 252102.
[36]    J. H. Chen, C. Jang, S. Adam, M. S. Fuhrer, E. D. Williams, M. Ishigami, *Nat. Phys.* **2008**, 4, 377.
[37]    J. Yan, M. S. Fuhrer, *Phys. Rev. Lett.* **2011**, 107, 206601.
[38]    K. Nomura, A. H. MacDonald, *Phys. Rev. Lett.* **2007**, 98, 076602.
[39]    A. W. Tsen, L. Brown, M. P. Levendorf, F. Ghahari, P. Y. Huang, R. W. Havener, C. S. Ruiz-Vargas, D. A. Muller, P. Kim, J. Park, *Science* **2012**, 336, 1143.
[40]    O. V. Yazyev, S. G. Louie, *Nat. Mater.* **2010**, 9, 806.
[41]    B. E. A. Saleh, M. C. Teich, *Fundamentals of Photonics*, Wiley Interscience, Hoboken, N.J. **2007**.
[42]    S. Bae, H. Kim, Y. Lee, X. Xu, J.-S. Park, Y. Zheng, J. Balakrishnan, T. Lei, H. Ri Kim, Y. I. Song, Y.-J. Kim, K. S. Kim, B. Ozyilmaz, J.-H. Ahn, B. H. Hong, S. Iijima, *Nat. Nanotechnol.* **2010**, 5, 574.
[43]    A. C. Ferrari, J. C. Meyer, V. Scardaci, C. Casiraghi, M. Lazzeri, F. Mauri, S. Piscanec, D. Jiang, K. S. Novoselov, S. Roth, A. K. Geim, *Phys. Rev. Lett.* **2006**, 97, 187401.

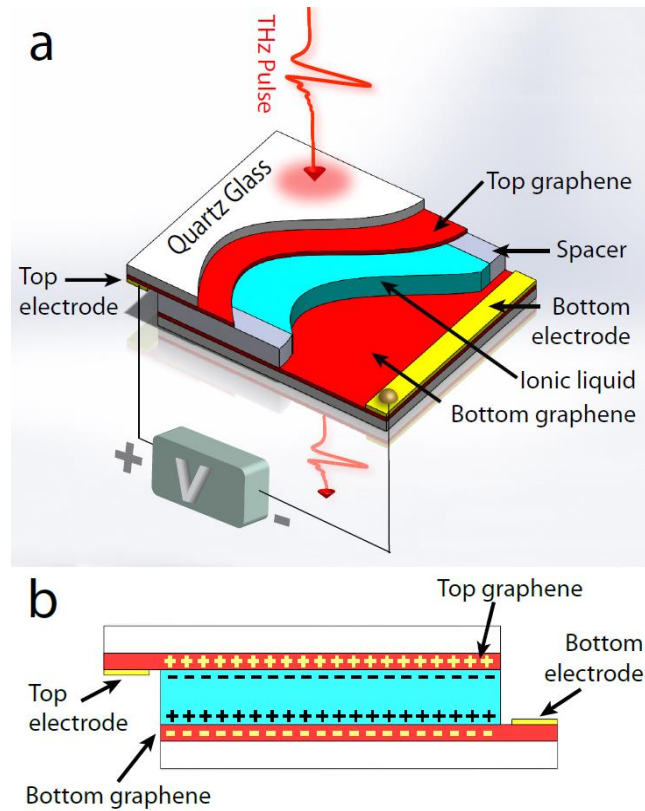

**Figure 1.** (a) The geometry of a graphene based THz modulator: the sandwich structure consists of two quartz glasses with graphene on top, two spacers and ionic liquid. The spacers are located at the edges of quartz glasses for supporting the cavity for ionic liquid. Cr/Au electrodes are deposited at one edge of each graphene film. (b) When voltages are applied as shown in (a), positive and negative charges accumulate at the graphene/ionic liquid interfaces.

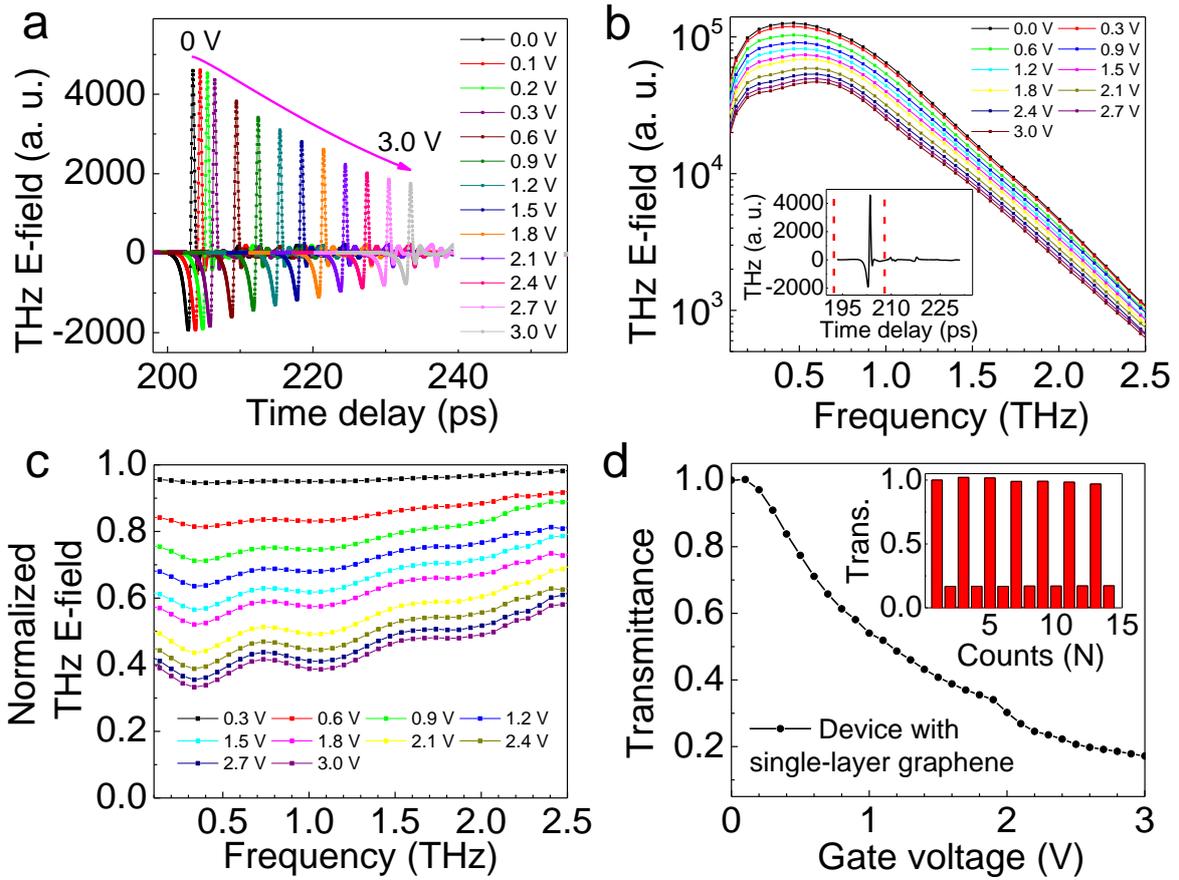

**Figure 2.** (a) Time domain electric field of THz pulses with the gate voltage up to 3 V. The THz spectra peaks decrease monotonically as a function of gate voltage. The data are shifted horizontally for clarity. (b) THz electric field amplitude in the frequency domain by the FFT. For FFT, time domain data are truncated between two vertical red lines (193.5 to 208 ps) to eliminate the multiple reflections (inset). (c) Normalized electric field strength with respect to the spectrum with zero gated voltage. (d) THz transmittance by averaging the normalized power intensity from 0.1 to 2.5 THz, which shows the maximum modulation of 83% with the gate voltage of 3 V. A repeatability test is performed by switching the gate voltage among 0 and ±3 V (inset).

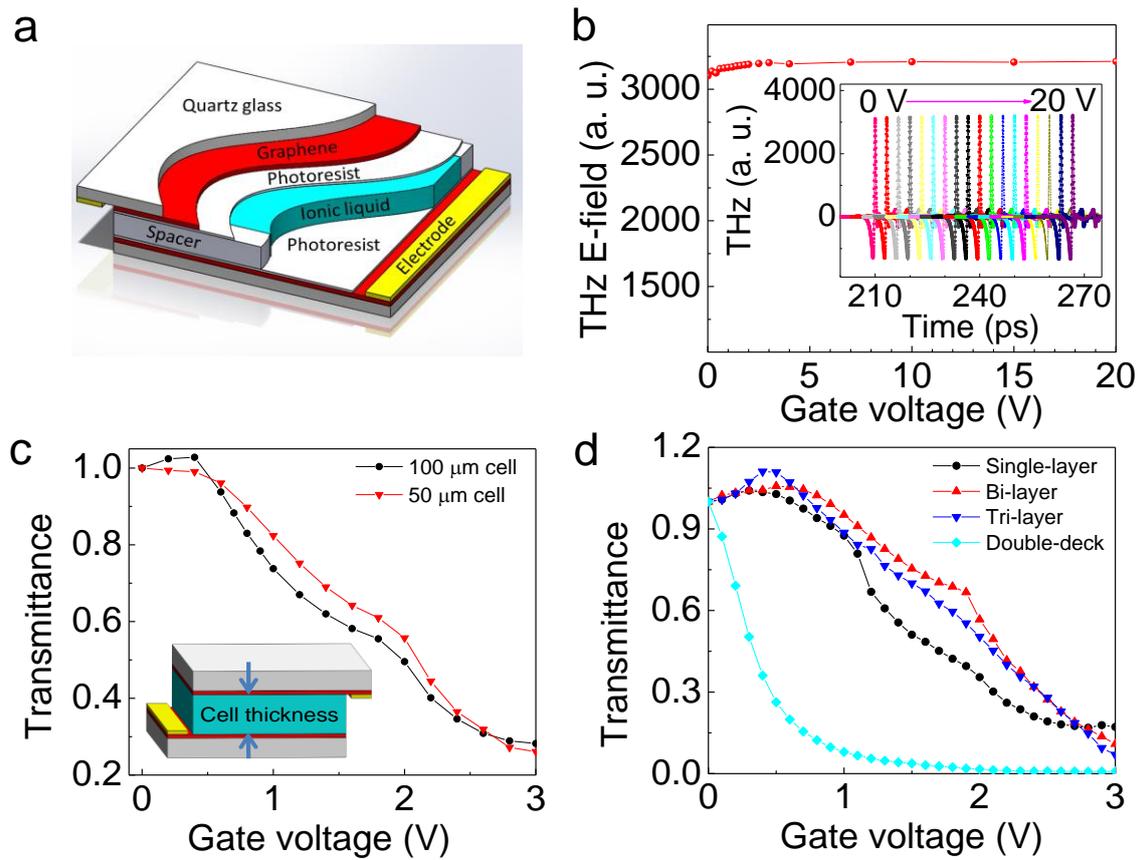

**Figure 3.** (a) Device structure for a control study; a thin layer of photoresist is coated on graphene. (b) Averaged THz signal from 0.1 to 2.5 THz at various gate biases. The inset is the time domain data with the curves shifted for clarify. (c) Normalized THz transmittance signal from devices with cell thickness of 50 and 100 μm. The inset is a schematic of device structure. (d) The comparison of THz modulation from devices with different numbers of graphene layer (single-layer, bi-layer, and tri-layer) showing a similar modulation depth. The double-deck device is fabricated by stacking a bi-layer and a tri-layer graphene device to enhance the modulation.
13

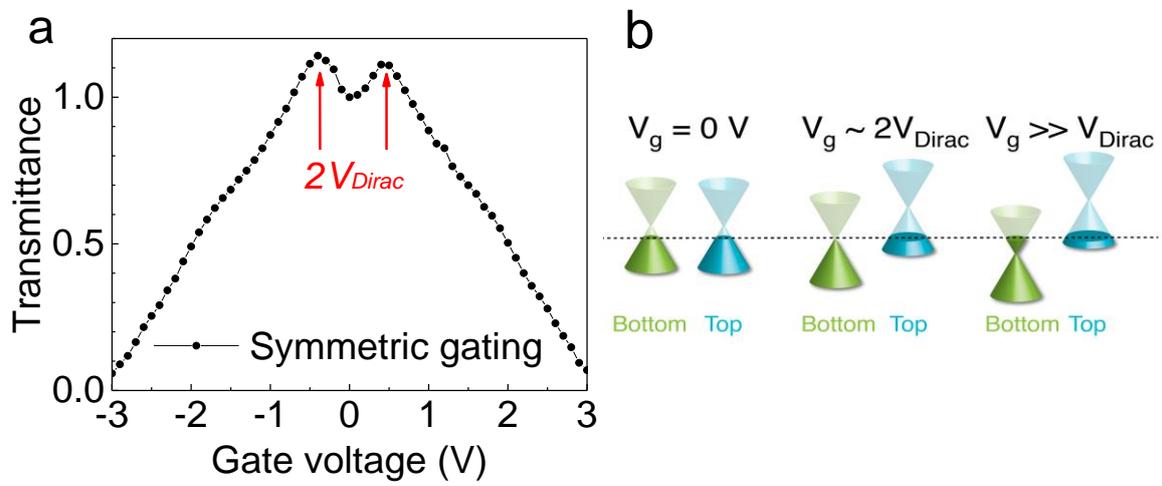

**Figure 4.** (a) Symmetric gate voltage modulation response of a tri-layer graphene device. (b) Band structures of top (blue) and bottom (green) graphene films under different gating conditions at $V_g = 0$, $V_g \approx 2V_{Dirac}$, and $V_g \gg V_{Dirac}$.

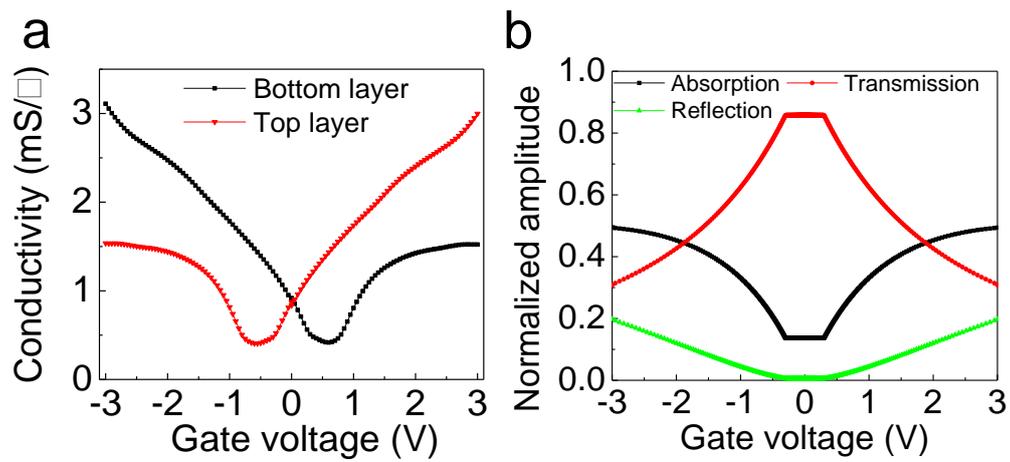

**Figure 5.** (a) Conductivity of the top and bottom graphene layers. (b) Simulated transmittance (red), absorption (black) and reflection (green) as a function of the gate voltage.

# Supporting Information

**Graphene terahertz modulators by ionic liquid gating**

*Yang Wu, Chan La-o-vorakiat, Xuepeng Qiu, Jingbo Liu, Praveen Deorani, Karan Banerjee, Jaesung Son, Yuanfu Chen, Elbert E. M. Chia,\* and Hyunsoo Yang\**

**1. Multiple reflections in the graphene cavity**

Multi-reflections are the phenomena that the incident wave experiences several reflections from internal interfaces, and then transmitting through the structure to show a much smaller peak compared to the main pulse. For any multi-reflection, due to the extra optical path, there is a time delay $\Delta t = 2dn/c$, where $d$ is the distance between two interfaces, $n$ is the refractive index (1 for dry air, 1.57 for ionic liquids, and 1.96 for quartz glasses at THz range), and $c$ is the speed of light. The schematic of one such possible multi-reflection is sketched in Fig. S1. As the thickness of ionic liquid cell and quartz glass is 100 and 400 μm respectively, the first order multi-reflections have a time delay of 1.05, 5.23, 6.28 ps, etc. In our data analysis, as shown in the inset of Figure 2b, only a small range (193.5 to 208 ps) was chosen so that the effect of multi-reflections are eliminated. Note that the first multi-reflection signal shown in Fig. S1 was still included in this range, because it is very close to the main pulse (1.05 ps delay), however, higher orders multi-reflection from these two graphene layers are neglected due to the much lower intensity.

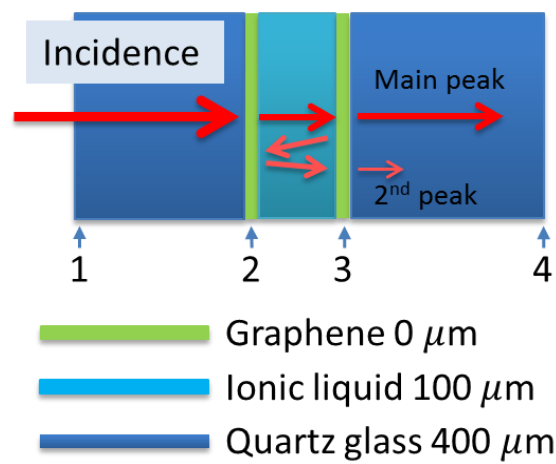

Fig. S1. Schematic of one multi-reflection of THz beam inside the devices.

## 2. Symmetric voltage response of THz modulation

The THz transmittance as a function of gate voltage in Figure 4 (main text) was processed starting from THz pulses in time domain (Fig. S2a). With a truncation from 193.5 to 208 ps, FFT was performed. Then, all the spectra were normalized to the spectrum with zero gate voltage (Fig. S2b). Finally, we average the transmittance over the measured frequency range to get the data in Figure 4a.

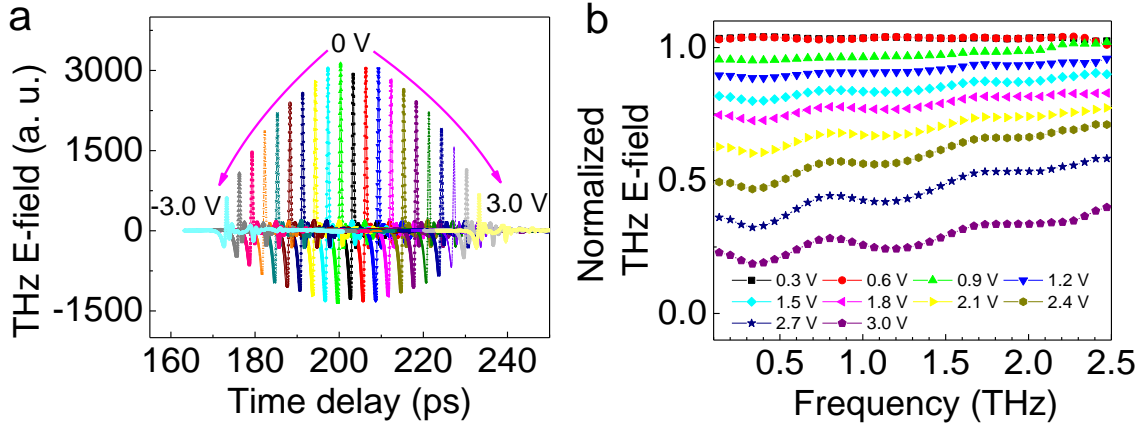

Fig. S2. Detail analysis for the tri-layer graphene based devices in Figure 4a. (a) Time domain spectroscopy of the devices under different gate voltage. The pulses are shifted horizontally for clarity. (b) Normalized THz electric field spectra after FFT.

## 3. Device simulation based on transfer matrix theory

According to the transfer matrix theory of a multilayer system, the propagation of light through an interface between two materials can be described by

$$S_{m1/m2} = \begin{bmatrix} t_{m1/m2} & r_{m2/m1} \\ r_{m1/m2} & t_{m2/m1} \end{bmatrix} = \begin{bmatrix} \dfrac{2n_{m1}}{n_{m1}+n_{m2}} & \dfrac{n_{m2}-n_{m1}}{n_{m1}+n_{m2}} \\ \dfrac{n_{m1}-n_{m2}}{n_{m1}+n_{m2}} & \dfrac{2n_{m2}}{n_{m1}+n_{m2}} \end{bmatrix} \quad \text{(Eq. 1)}$$

where $t$ and $r$ correspond to the transmission and reflection coefficients of the electric field, respectively. The subscription $m1$ stands for the material 1 and $m2$ for material 2. Graphene films are assumed to be zero thickness conductive layers with the vacuum impedance of $Z_0 = 376.73\ \Omega$. Then the transfer matrix for the structure of $m1$/graphene/$m2$ can be written as

$$S_{m1/graphene/m2} = \begin{bmatrix} t_{m1/graphene/m2} & r_{m2/graphene/m1} \\ r_{m1/geaphene/m2} & t_{m2/graphene/m1} \end{bmatrix} = \begin{bmatrix} \dfrac{2n_{m1}}{n_{m1}+n_{m2}+Z_0\sigma_s} & \dfrac{n_{m2}-n_{m1}-Z_0\sigma_s}{n_{m1}+n_{m2}+Z_0\sigma_s} \\ \dfrac{n_{m1}-n_{m2}-Z_0\sigma_s}{n_{m1}+n_{m2}+Z_0\sigma_s} & \dfrac{2n_{m2}}{n_{m1}+n_{m2}+Z_0\sigma_s} \end{bmatrix} \quad \text{(Eq. 2)}$$

where $\sigma_s$ is the conductivity of graphene.

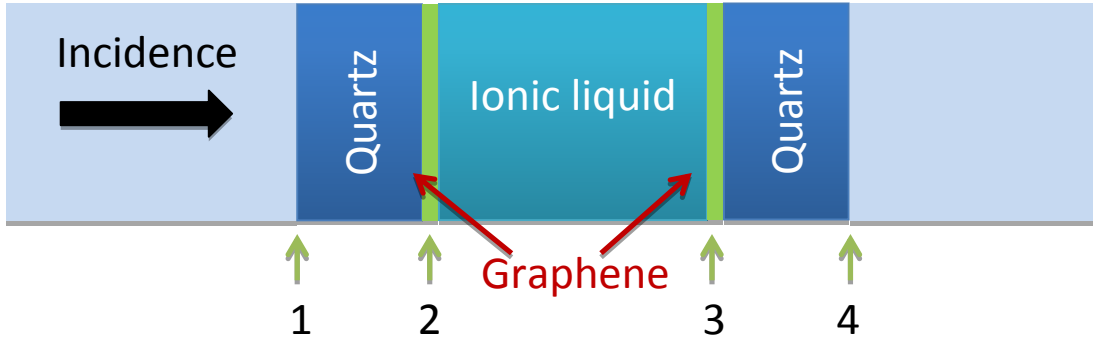

Fig. S3. Simulation model for typical sandwiched THz modulators consisted of dry air/quartz glass/graphene/ionic liquid/graphene/quartz glass/dry air.

Our THz devices are a stack of quartz glass/ionic liquid/quartz glass (Fig. S3). When transmitting THz wave through, four interfaces are involved for our investigation (labelled as 1-4 in Fig. S3). A single $M$ matrix is derived from the $S$ matrix in Eq. (1 - 2) and defined as

$$M_{m1/m2} = \frac{1}{t_{m2/m1}} \begin{bmatrix} t_{m1/m2}t_{m2/m1} - r_{m1/m2}r_{m2/m1} & r_{m2/m1} \\ -r_{m1/m2} & 1 \end{bmatrix} \quad \text{(Eq. 3)}$$

The transfer matrix describing the stack in Fig. S3 is

$$M = M_{air/quartz} \times M_{quartz/graphene/ionic\_liquid} \times M_{ionic\_liquid/graphene/quartz} \times M_{quartz/air}. \quad \text{(Eq. 4)}$$

With the $M$ matrix for the entire stack, we can convert this $M$ matrix back to the $S$ matrix from which the reflectance R = $r^2$, transmittance T = $t^2$, and absorption A = 1-R-T can be obtained. As the gate voltage changes the conductivity of the graphene layers in our device, the absorption, reflection, and transmittance are modulated accordingly. The calculated results are presented in Fig. S4. The $x$-axis is shifted by 0.3 mS to match the characteristic of $p$-doped graphene films. Using the measured conductivity at different gate biases, the absorption, transmission, and reflection as a function of gate bias are obtained as shown in Figure 5b.

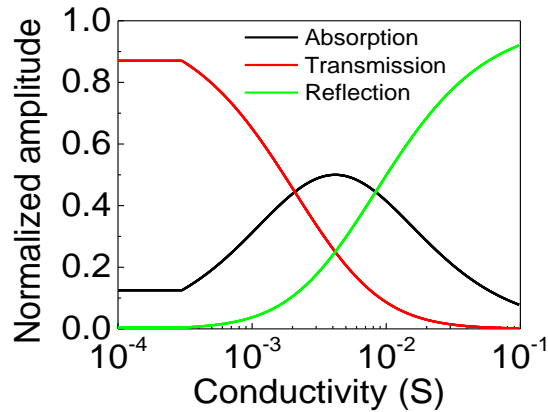

Fig. S4. Simulated absorption, transmission, and reflection as function of conductivity.

## 4. Terahertz time-domain spectrometer

Our THz time domain spectroscopy system in Fig. S5 is based on an ultrafast fs-laser. The laser beam is split into two parts for the THz generation and detection. The generation beam excites a biased LT-GaAs photoconductive antenna (PCA) to radiate THz waves. The THz spectra cover the range from 0.1 to 2.5 THz. After passing through the samples, the THz wave is focused to the other PCA for detection. Meanwhile, the detection beam travels through a delay line and incidents onto the detection PCA. When the THz pule and the detection beam reach the PCA at the same time, a transient current is generated, which is proportional to the THz electric field strength. By scanning the optical path differences between the detection beam and THz beam, time profiles of THz pulses can be constructed. The scanner is operated at frequency of 30 Hz to improve the signal-to-noise ratio. The entire beam path is purged with dry air to avoid the absorption from atmospheric water vapor.

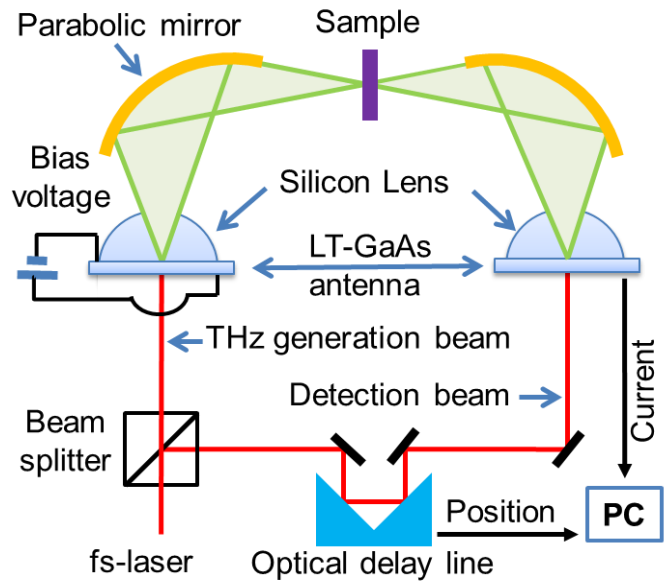

Fig. S5. Experimental setup of THz time-domain spectrometer.